# Near band-edge luminescence of semi-insulating undoped gallium arsenide at high levels of excitation

V.F. Kovalenko (1), S.V. Shutov (1,2), Ye.A. Baganov (1), M.M. Smyikalo (1)

(1) Kherson National Technical University, 24 Berislavskoye shosse, Kherson 73008, Ukraine
(2) V. Lashkarev Institute of Semiconductor Physics, National Academy of Sciences of Ukraine, 41 prospect Nauki, Kiev 03028, Ukraine.

**Abstract**
The dependences of the maximum and the half-width of near band-edge photoluminescence of semi-insulating undoped GaAs crystals at 77K on the concentration of background acceptor impurities and the level of excitation in the range from $3 \cdot 10^{21}$ to $6 \cdot 10^{22}$ quantum/(cm$^2$·s) are investigated. The observed dependences are explained by formation of the density tails of states as a result of fluctuations of impurity concentration and participation of localized states of the donor impurity band in radiative transitions. Reduction of many-particle interaction at increasing of N can be connected with increasing of shielding of charge carriers by atoms of impurity.

# 1. Introduction

Investigations of different kinds of many-body interactions in direct-gap semiconductors, including electron-hole plasma, have been carried out, as a rule, for materials with high concentration of impurity. But at high levels of excitation $L$ even in crystals with low impurity concentration effects of many-body interactions take place due to high concentration of charge carriers.

Widening of GaAs application to SHF devices maintains the interest in investigations of such phenomena [1-4]. Earlier [5] some characteristic properties of near band-edge (NBE) photoluminescence (PL) of semi-insulating undoped (SIU) GaAs crystals have been investigated at low levels of excitation ($L \leq 10^{21}$ quantum/(cm$^2$·s)). In [6, 7] we gave the results of investigations of behavior of near band-edge photoluminescence and conditions of appearing of electron-hole pairs in SUI GaAs crystals at different levels of excitation. Background impurity was carbon with the concentration in the range from $3 \cdot 10^{15}$ cm$^{-3}$ to $4 \cdot 10^{16}$ cm$^{-3}$.

In the present work we shall give the results of investigations of the NBE PL of the SUI GaAs crystals with Si and C background impurities at high levels of excitation ($L > 10^{21}$ quantum/(cm$^2$·s)). We shall show, that in spite of the fact that the results obtained by us and in [5] are close, for their explanation it is necessary to use other concepts.

# 2. Experimental technique

GaAs crystals grown by the Chokhralsky method with specific resistance $1 \cdot 10^7 \div 5 \cdot 10^8$ Ohm·cm and electronic conductivity due to the presence of EL2 defects were investigated. The concentration $N$ of the basic background acceptor impurity (carbon and silicon) was in the range of $3.5 \cdot 10^{15} \leq N \leq 4 \cdot 10^{16}$ cm$^{-3}$. Estimation of $N$ was carried out by the method, described in [8]. The investigated crystals were characterized by the average density of the dislocations in the range from $3.2 \cdot 10^4$ to $1.5 \cdot 10^5$ cm$^{-2}$. Measurements of the NBE PL characteristics (the energy of a maximum $h\upsilon_m$, the half-width $W$, and the integrated intensity $I$) were carried out accordingly to [9] using of excitation radiation of argon laser ($\lambda_b$=0.488 – 0.514 μm) within the intensity limits $L$ in the range $3 \cdot 10^{21} \leq L \leq 6 \cdot 10^{22}$ quantum/(cm$^2$·s) at temperature 77 K. The precision of measurements of $h\upsilon_m$ and $W$ was not lower than 0.5 meV and 0.6 meV accordingly.

# 3. Results and discussion

*3.1. Dependences on concentration of background impurity.* The short-wave range of PL spectrum contained a NBE radiation and additional longer wave (15-17 meV distant) band of low intensity, which is connected with participation of centers $C_{As}$ and $Si_{Ga}$ in radiative transitions [5]. The latter band will not be taken in consideration.

In fig. 1 (cur. 1) the dependence of $W$ of near-edge band of PL on $N$ at $L = 3 \cdot 10^{21}$ quantum/(cm$^2$·s) are represented. These data indicate the monotonous W increasing from 6 – 6.2 meV in crystals with the lowest values of $N$ up to 16.5 meV in crystals with the maximal values of concentration of background acceptor impurity due to the widening of the near-edge band both to low-energy and to high-energy areas of a spectrum. Thus $h\upsilon_m$ remains practically constant at $N < 1.5 \cdot 10^{16}$ cm$^{-3}$ and decreases with increasing of the concentration of an impurity at $N > 1.5 \cdot 10^{16}$ cm$^{-3}$ (fig. 1, cur.2). The increasing of impurity concentration results in increasing of $I$ in a whole investigated range of $N$ (fig. 1, cur.3).

**3.2. Dependences PL on the level of excitation.** Dependences of $W$ of the near-edge band and $h\upsilon_m$ of PL in crystals with different values of $N$ on $L$ are shown in the fig. 2. Increasing of $L$ in the investigated range resulted in the expansion of a band accompanied with the shift of $h\upsilon_m$ to the low-energy area of a spectrum. The most essential reorganization of a spectrum at increasing of $L$ was observed in crystals with the lowest values of $N$. Influence of $L$ on PL spectrum decreased with increasing of $N$. Increase of $L$ results in increase of $I$ of NBE PL as the second-degree function.

**3.3. Discussion of results.** Estimations show [10] that in similar crystals and at comparable $L$ applied in present work the concentration of minority carriers exceeds $10^{17}$ cm$^{-3}$. At such concentration there is no formation of exciton [11], because the difference $E_g - h\upsilon_m$ (where $E_g$ is the bandgap) is less than the corresponding binding energy of an exciton [12]. Therefore the observed dependences $h\upsilon_m = f(L)$ and $W = f(N, L)$ cannot be explained in the way of annihilation of free and bound excitons being in the ground state or in the first excited state. On the other hand, the width of the PL band in the purest crystals (about $kT$) is the feature of the excitonic radiating recombination. Thereupon the following explanation of obtained dependences is proposed.

According to [13] in the direct-gap semiconductors at band-to-band radiative transitions of non-degenerate electrons and holes without taking into account their interaction and dispersion the theoretical value of the width of the PL band is equal to 1.8 $kT$, and the $h\upsilon_m$ is close to $E_g$. When Coulomb interactions are presence a luminescence should be considered as a result of annihilation of excitons, being in the state of a continuous spectrum. Interaction of charge carriers at recombination thus results in both reduction of $W$ up to 0.7 $kT$ and shift value of $h\upsilon_m$ to the long-wave area on value, which is smaller than binding energy of free exciton. Dispersion of recombining charge carriers stipulates the contribution of indirect transitions to PL and at presence of Coulomb interactions gives the values of $W$ in the range from 0.7 $kT$ to 0.9 $kT$.

Observed values of $W$ in the crystals with $N \leq 1.5 \cdot 10^{16}$ cm$^{-3}$ and $h\upsilon_m = 1.5096$ eV which are in accordance with theoretical ones allow to assume the presence of interaction between recombining free carriers at indirect transitions. Expansion of the near-edge band to a short-wave range of a spectrum caused by increase of $N$ proves the contribution of dispersion into PL spectra and, hence, increasing the fraction of indirect transitions.

Dependence of spectral characteristics of NBE PL on impurity concentration can be connected with degradation of edges of the valence and the conduction bands, caused by local fluctuations of total concentration $N_\Sigma$ of electrically active impurity and defects ($C_{As}$, EL2, $Si_{Ga}$, etc.). Significant excess of value of $N_\Sigma$ (several orders greater) over the concentration of free electrons $n_0$ makes practically impossible the electron shielding of spatial charges. We assume, that this fact provides a formation of potential wells (with the width $r$ and the energy depth $\gamma$) which results, despite of low concentration $N_\Sigma$, in essential degradation of edges of the valence and the conduction bands in crystals SIU GaAs and being in contrast with the low-resistance GaAs crystals with the same concentration of an impurity [13].

Estimation of $\gamma$ can be made with assumption that the recombination of non-localized charge carriers occurs from the levels of passing $\gamma_{e,h}$ [13], which energy position from the edges of corresponding bands is $\gamma_{e,h} \cong \frac{2}{3}\gamma$ [13]. Therefore the maximal value of $\gamma$ in crystals with the largest contents of the background acceptors $N$ can be estimated from the expression:

$$\gamma \cong \frac{3}{4}(E_g - h\nu_m)$$

The value of $\gamma$ is equal to 0.0045 eV. The well width $r$ can be determined from the formula [13]:

$$\gamma = \sqrt{2\pi}\frac{e^2}{\varepsilon}\sqrt{N_\Sigma r}$$

where $e$ is the charge of the electron, $\varepsilon$ is the permittivity of GaAs. In the assumption, that $N_\Sigma$ does not exceed $6\cdot 10^{16}$ cm$^{-3}$ (the maximal impurity concentration is $4\cdot 10^{16}$ cm$^{-3}$, concentration of defects EL2 is $2\cdot 10^{16}$ cm$^{-3}$ [5]), one can obtain $r = 6\cdot$nm. Such parameters of tails of state do not provide a condition of carrier localization $(\hbar^2/m^*_{e,h}r) < \gamma$, where $\hbar$ is the Planck constant, $m^*_{e,h}$ is the effective mass of an electron or a hole.

In spite of absence of localization of charge carriers in tails of state, the influence of these tails of state is, in our opinion, observed in radiation transitions between states close to levels of passing which is shifted into the bandgap (fig. 3). The density of these states is lower, than in the valence and the conduction bands [12]. It provides widening of PL band with increasing of $N$. Simultaneous decreasing of $h\nu_m$ can be explained by increasing of shift of levels of passing as a result of increasing of well depth [12].

The shift of $h\nu_m$ to the long-wave range and increasing of $W$ with increasing of $L$ are the characteristic attributes of formation of electron-hole plasma and its radiative recombination [7, 14]. The estimation of concentration of nonequilibrium electrons and holes $\delta_n$ and $\delta_p$ respectively as well as density of electron-hole plasma $n_{e,h}$, was carried out by the formula:

$$\delta_n = \delta_p = n_{e,h} = \beta L'\alpha\tau$$

where $\beta = 1$ is the quantum yield of an intrinsic photoeffect, $L' = (1-R)L$ is the intensity of the absorbed exciting radiation, $R \cong 0.3$ is the factor of reflection, $\alpha \approx 3.5 \div 4\cdot 10^4$ cm$^{-1}$ is the factor of absorption of exciting radiation, $\tau \approx 250$ picoseconds [15] is the life time of nonequilibrium carriers. In assumption that $\tau$ does not depend on $L$ it can be obtained that the value of $n_{e,h}$ is in the range $1.9\cdot 10^{16}$ cm$^{-3} \leq n_{e,h} \leq 3.7\cdot 10^{17}$ cm$^{-3}$ at $L$ changes from $3\cdot 10^{21}$ to $6\cdot 10^{22}$ quantum/(cm$^2\cdot$s). The obtained range of values $n_{e,h}$ includes the value $n_{e,h} = 2 \div 4\cdot 10^{16}$ cm$^{-3}$ [14], corresponding to equilibrium density of electron-hole plasma in GaAs. This fact confirms the assumption that the main reason of reorganization of PL spectrum with increasing of *L* is the formation of electron-hole plasma and its radiative recombination.

## 4. Conclusions

Investigations of NBE PL of SIU-GaAs crystals with carbon and silicon as background acceptor impurity at high levels of excitation allow to suppose, that:

1. Dependence of both energy of the maximum and the half-width of near-edge band of PL on concentration of background impurity is caused by influence of the density tails of states appeared due to fluctuation of concentration of impurity. Band edge tailing occurs due to participation of the states that are localized near the levels of passing and shifted into the bandgap in radiative transitions;

2. Reduction of many-body interaction with increasing of the impurity concentration can be connected with increase of shielding of interacting carriers by the ionized atoms of impurity and defects;

3. Dependence of both energy of the maximum and the half-width of near-edge band of PL on the level of excitation is caused by interactions between charge carriers that is formation and radiative recombination of the electron-hole plasma;

4. At high levels of excitation the shape of near-edge band of PL is determined by recombination of free interacting charge carriers at presence of indirect transitions.

**Acknowledgements**

Authors thank K. D. Glinchuk for useful and substantial discussion of experimental results.

**Figure captions**

Fig. 1. Dependences of the half-width $W$ (1), the energy of a maximum $h\upsilon_m$ (2), and the integrated intensity $I$ (3) of the NBE PL on the impurity concentration $N$. Level of excitation $L$ is $3 \cdot 10^{21}$ quantum/(cm$^2 \cdot$s).

Fig. 2. Dependences of the half-width $W$ (curves 1-3) and the energy of a maximum $h\upsilon_m$ (curves 1'-3') of NBE PL on the level of excitation $L$. Concentration of impurity (in cm$^{-3}$): 1, 1' – $3.5 \cdot 10^{15}$; 2, 2' – $1.3 \cdot 10^{16}$; 3, 3' – $2.1 \cdot 10^{16}$.

Fig. 3. Energy diagram and the channel of radiative recombination in SIU-GaAs with the increased concentration of background easy ionizable impurity. $E_{Co}$, $E_{Vo}$ are the edges of bands of the "not excited" semiconductor; $E_C$, $E_V$ are the levels of passing for electrons and holes respectively of the semiconductor. Hatching denotes the allowed states of bands.

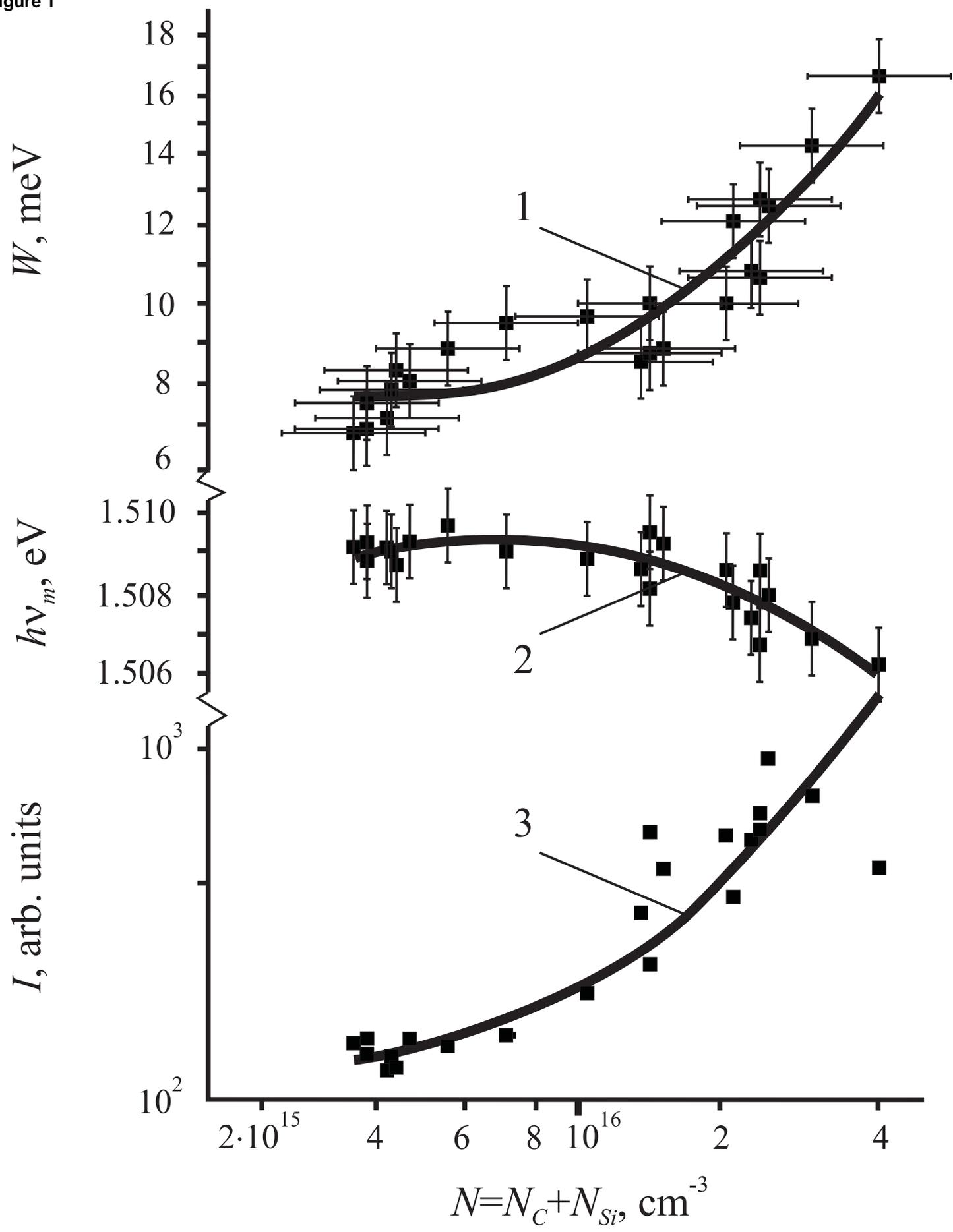

Figure 1

Figure 2

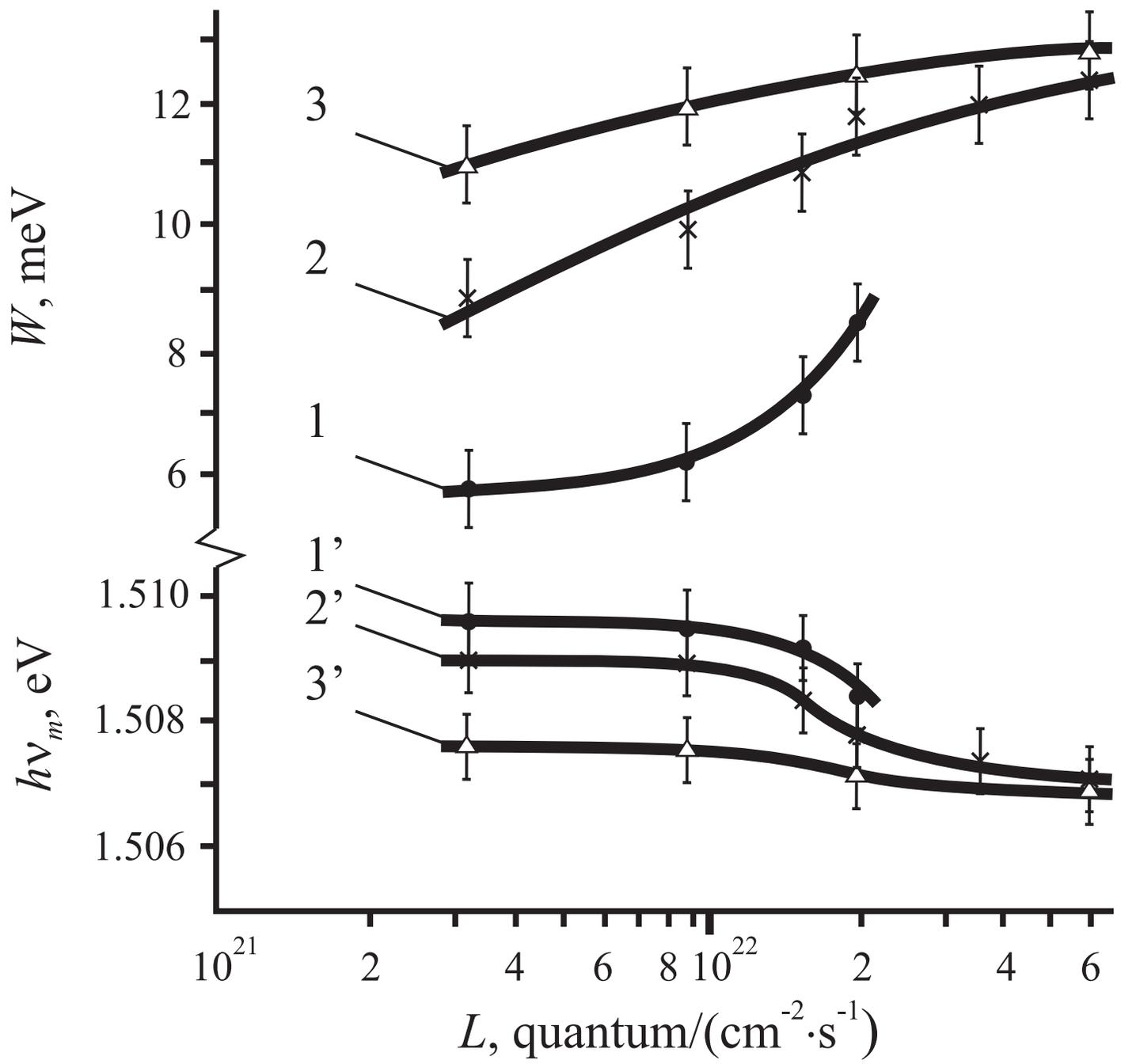

**Figure 3**

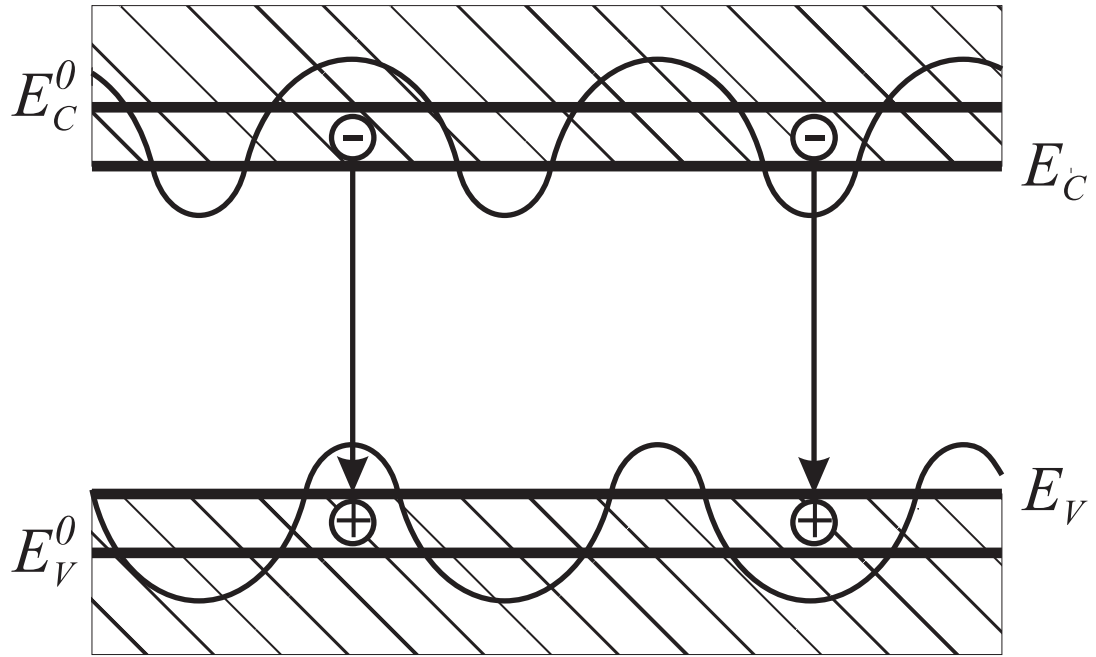